\documentclass[reprint,showpacs,
superscriptaddress,
 amsmath,amssymb,
 aps,prl
]{revtex4-1}
\usepackage{amsmath}
\usepackage{amsfonts}
\usepackage{amssymb}
\usepackage{graphicx}

\begin{document}

\title{Flow reversals in turbulent convection via vortex reconnections}

\author{Mani Chandra}
\author{Mahendra K. Verma}
\email{mkv@iitk.ac.in}
\affiliation{Department of Physics, Indian Institute of Technology - Kanpur 208016, India}

\begin{abstract}
We employ detailed numerical simulations to probe the mechanism of flow reversals in two-dimensional turbulent convection.  We show that the reversals occur via vortex reconnection of two attracting corner rolls having same sign of vorticity, thus leading to major restructuring of the flow.   Large fluctuations in heat transport are observed during the reversal due to this flow reconfiguration.  The flow configurations during the reversals have been analyzed quantitatively using large-scale modes.  Using these tools, we also show why flow reversals occur for a  restricted range of Rayleigh and Prandt numbers.  
\end{abstract}


\maketitle
Several experiments~\cite{Sugiyama:PRL2010,Cioni:JFM1997, Niemela:JFM2001,
Brown:JFM2006, Xi:PRE2007, Yanagisawa:PRE2010, Gallet:GAFD2012, Vasiliev:JCONF2011} and numerical
simulations~
\cite{Sugiyama:PRL2010,Benzi:EPL2008, Breuer:EPL2009,Mishra:JFM2011, Chandra:PRE2011} on
turbulent convection exhibit ``flow reversals'' in which  the probes near the lateral walls of the container show random reversals (also see review articles~\cite{Ahlers:RMP2009}).  These reversals have certain similarities with magnetic field reversals in dynamo and Kolmogorov flow~\cite{Gallet:GAFD2012}.   Researchers typically study convection in a controlled setup called
``Rayleigh-B\'{e}nard convection" in which a fluid confined between two plates
is heated from below and cooled at the top.  The two nondimensional numbers used
to characterize the flow are the Rayleigh number ($\mathrm{Ra}$), which is the ratio of
the buoyancy term and the diffusive term, and the Prandtl number ($\mathrm{Pr}$), which
is the ratio of the kinematic viscosity and the thermal diffusivity.   Cioni {\em et al.}~\cite{Cioni:JFM1997}, Niemela
{\em et al.}~\cite{Niemela:JFM2001}, Brown and Ahlers~\cite{Brown:JFM2006}, and
Xi {\em et al.}~\cite{Xi:PRE2007} performed convection experiments on mercury, water, and
helium gas in a cylindrical geometry and observed reversals for
$\mathrm{Ra}>10^8$.  Sugiyama et al.~\cite{Sugiyama:PRL2010} and Vasiliev and
Frick \cite{Vasiliev:JCONF2011} studied reversals in a rectangular box with water.  No reversal was observed for a cubical box (aspect ratio 1), but a quasi two-dimensional box (aspect ratio $\le 0.2$) exhibits reversals for a band of Rayleigh and Prandtl numbers~\cite{Sugiyama:PRL2010}.   Surprisingly, a cubical box containing mercury shows reversals~\cite{Gallet:GAFD2012}, indicating a strong role played by the Prandtl number and geometry in the reversal dynamics.

Several theoretical models have been invoked to explain flow reversals.  Benzi and Verzicco~\cite{Benzi:EPL2008} and Sreenivasan {\em et al.}~\cite{Sreenivasan:PRE2002} used stochastic resonance, while Arajuo {\em et al.}~\cite{Araujo:PRL2005} employed low-dimensional models with noise to explain reversals.  Brown and Ahlers~\cite{Brown:JFM2006} and Mishra {\em et al.}~\cite{Mishra:JFM2011} showed that in a cylindrical geometry, the flow reversals are induced by a rotation or cessation of large-scale flow structures.   For two-dimensional box geometry, Sugiyama {\em et al.}~\cite{Sugiyama:PRL2010} relate the flow reversal to the growth of the corner rolls due to the plume detachments from the boundary layers.  Chandra and Verma~\cite{Chandra:PRE2011}  studied the reversals quantitatively by representing flow structures as Fourier modes and showed that during the reversals, the amplitude of the first Fourier mode $(k_x=1,k_y=1)$ becomes very small, while the Fourier mode $(k_x=2,k_y=2)$ gains strength.    The growth of the secondary modes at the expense of the primary modes is akin to the cessation-led reversals reported by Brown and Ahlers~\cite{Brown:JFM2006} and  Mishra {\em et al.}~\cite{Mishra:JFM2011}, and to the emergence of quadrupolar mode in  dynamo reversals~\cite{Gallet:GAFD2012}.  

One important question that persists is why do flow reversals occur in a
restricted parameter regime? In this letter, we explain this through a quantitative
investigation of the convective flow structure for a range of parameters using Fourier
basis decomposition.  This scheme allows for accurate representation of the flow~\cite{Chandra:PRE2011} and lets us quantify when reversals occur in terms of amplitudes of the modes.  In addition we show that flow reversals occur via vortex reconnections, and connect them to vortex dynamics.   We
investigate in detail the heat transport in the system during
a reversal and find large fluctuations in it as a result of the flow reorganization
during the reversals. We restrict our study to 2D flow reversals whose flow
structures are well represented by experiments conducted in quasi two-dimensional
geometry~\cite{Sugiyama:PRL2010}.

%The equations governing Rayleigh-B\'enard convection under Boussinesq approximation are
%\begin{eqnarray}
%    \frac{\partial \mathbf{u} }{\partial t} + (\mathbf{u} \cdot \nabla)
%    \mathbf{u} & = & -\nabla P + \mathrm{Pr} \nabla^{2} \mathbf{u} + \mathrm{Ra} \mathrm{Pr} T
%    \hat{y} \label{eqn:vel_eqn}\\ \frac{\partial T}{\partial t} +
%    (\mathbf{u} \cdot \nabla) T & = & \nabla^{2} T \\ \label{eqn:temp_eqn}
%    \nabla \cdot \mathbf{u} & = & 0 \label{eqn:div_free_condition}
%\end{eqnarray}
%where $\mathbf{u}$ is the velocity field, $T$ is the temperature field, $P$ is the pressure, and $\hat{y}$ is the buoyancy direction. The nondimensional parameters that appear in the equations are (1) the Prandtl number $\mathrm{Pr} = \nu/\kappa$, where $\nu$ is the kinematic viscosity and $\kappa$ is the thermal diffusivity, and (2) the Rayleigh number $\mathrm{Ra} = \alpha g \Delta d^3/(\nu \kappa)$, where $\alpha$ is the thermal expansion coefficient, $d$ is the distance between the two plates, $\Delta$ is the temperature difference between the plates, and $g$ is the acceleration due to gravity. The equations have been ondimensionalized taking $d$ as the length scale, and the diffusion time $d^2/\kappa$ as the time scale.

\begin{figure*}
    \begin{center}
        \includegraphics[scale=0.25]{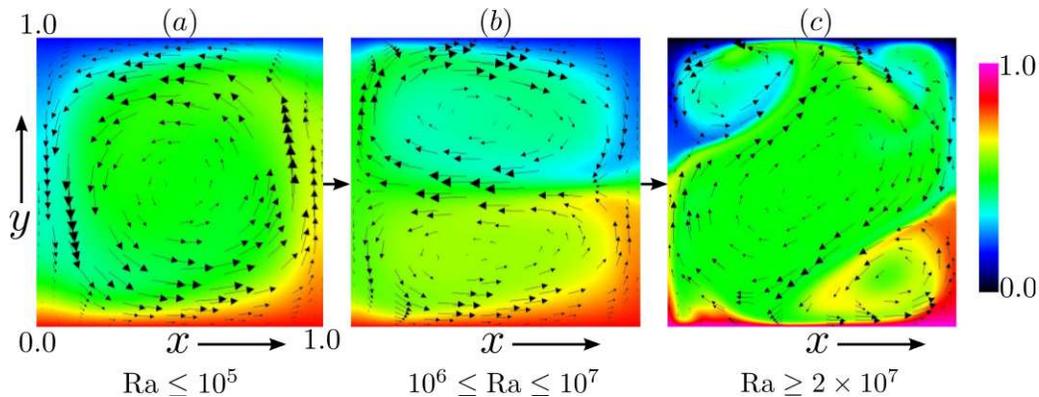}
    \end{center}
    \caption{Steady state flow structures at different Rayleigh numbers: (a) for $Ra \le  10^5$, a single roll with the dominant $(m=1,n=1)$ mode, (b) for $10^6 \le Ra \le 10^7$, two rolls stacked on top of each other corresponding to the mode
    $(1,2)$,  and (c) for $Ra \ge 2 \times 10^7$, two corner rolls with a dominant roll aligned along the 45 degree diagonal, a configuration dominated by the modes $(1,1)$ and     $(2,2)$. }
    \label{fig:phase_transformation}
\end{figure*}

%We solve equations (\ref{eqn:vel_eqn}-\ref{eqn:div_free_condition})  in a 2D box
We solve the equations governing Rayleigh-B\'{e}nard convection under Boussinesq approximation~\cite{Chandra:PRE2011}  in a 2D box of aspect ratio 1 with no-slip walls on all sides.  The side walls are insulating, while  the top and the bottom walls are maintained at constant temperatures.  The simulations are performed using the open-source code NEK5000
\cite{Fischer:JCP1997} that employs spectral element method.  We use $28\times28$ spectral element with $7^{th}$ order polynomial, thus we have an effective grid of $196\times196$ points. The grid density is higher at the boundaries in order to resolve the boundary layers.   We perform simulations for $\mathrm{Pr} =1 $ and Rayleigh numbers ranging from $10^5$ to $10^9$. 

To quantify the flow structures, we project the nondimensionalized horizontal velocity ($u$), vertical velocity ($v$),  and temperature ($T$) onto the following basis:
\begin{eqnarray}
 u & = & \sum_{m,n}  \hat{u}_{m,n}  \sin(m \pi x) \cos(n \pi y) \\
 (v,T) & = & \sum_{m,n} (\hat{v}_{m,n},  \hat{T}_{m,n}) \cos(m \pi x) \sin(n \pi y) 
 \end{eqnarray}
It has been shown that the above basis is a good representative of the flow field in a box geometry~~\cite{Chandra:PRE2011}.
We choose the above formalism over proper orthogonal decomposition (POD) in order to study the temporal evolution of the flow structures.  The mode with wavenumber $(m,n)$ corresponds to
a flow structure with $m$ rolls in the $x$ direction and $n$ rolls in the $y$
direction. In the following discussion we will use the above basis to analyze the mechanism of flow reversals as well as the range of $\mathrm{Pr}$ and $\mathrm{Ra}$ for which reversals take place.

An important puzzle in 2D turbulent convection is why flow reversals take place only for a range of Prandtl and Rayleigh numbers.  It has been shown that the corner rolls (represented by $(m=2,n=2)$  mode) are crucial for the occurrence of reversals~\cite{Sugiyama:PRL2010,Chandra:PRE2011}.  In particular, the relative strength of the (2,2) mode with respect to the dominant mode determines if a reversal will occur or not.   We observe three distinct flow structures for the range of $\mathrm{Ra}=(10^5 -10^9)$ performed in our simulations (see Fig.~\ref{fig:phase_transformation}).  The  mode (1,1), representing the large single role, is dominant till  $\mathrm{Ra} \sim 10^5$ (Fig.~\ref{fig:phase_transformation}(a)).  After this, the  mode (1,2) representing the two horizontally stacked rolls dominates till $\mathrm{Ra} \approx 10^7$ (Fig.~\ref{fig:phase_transformation}(b)).  The  mode (2,2), which is a dominant player for the flow reversals, is born only after $\mathrm{Ra} \approx 2\times  10^7$ (Fig.~\ref{fig:phase_transformation}(c)), which  is the reason for the absence of reversals for lower $\mathrm{Ra}$.   

\begin{figure}
    \begin{center}
        \includegraphics[scale=0.4]{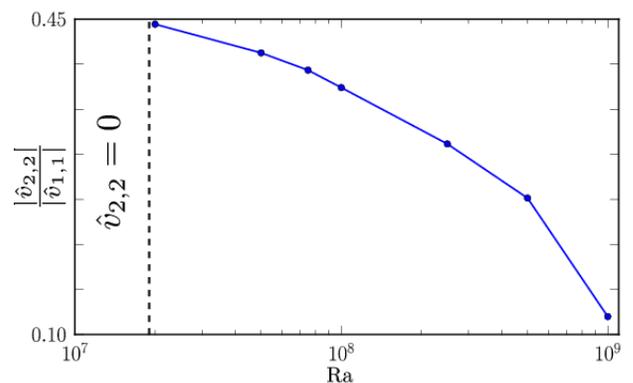}
    \end{center}
    \caption{Plot of the ratio $|\hat{v}_{2,2}|/|\hat{v}_{1,1}|$  vs.~$\mathrm{Ra}$.  The mode $\hat{v}_{2,2}$, which quantifies the strength of the corner rolls, is born only near $\mathrm{Ra} = 2\times 10^7$.  The average value of the ratio $|\hat{v}_{2,2}| / |\hat{v}_{1,1}|$ decreases from 0.45 to 0.10 as $\mathrm{Ra}$ increases from $2\times 10^7$ to $10^9$.   Reversals are observed for a narrow band near $\mathrm{Ra} = 2\times 10^7$ where $\hat{v}_{2,2}$ is strong.}
    \label{fig:mode_ratio}
\end{figure}

In the third regime, $\mathrm{Ra} \ge 2\times 10^7$, the averaged value of $|\hat{v}_{2,2}| / |\hat{v}_{1,1}|$ falls monodically from 0.45 at $\mathrm{Ra}=2 \times 10^7$ to 0.10  at $\mathrm{Ra}=10^9$ (see Fig.~\ref{fig:mode_ratio}).   The decrease in $|\hat{v}_{2,2}| / |\hat{v}_{1,1}|$ is due to the increase of $|\hat{v}_{1,1}|$,  possibly due to inverse cascade of energy.   This result indicates that for higher $\mathrm{Ra}$, the corner rolls become weaker compared to the (1,1) mode.  Therefore, reversals are observed only for a narrow band near $\mathrm{Ra}=2 \times 10^7$ where the (2,2) mode is strong enough. We observe similar behavior for $\mathrm{Pr}=10$ except that the (2,2) mode appears for $\mathrm{Ra}\approx 10^6$, a value lower than that for $\mathrm{Pr}=1$.  Consequently the range of $\mathrm{Ra}$ exhibiting flow reversals is broader for $\mathrm{Pr}=10$ compared that $\mathrm{Pr}=1$, consistent with the results of Sugiyama {\em et al.}~\cite{Sugiyama:PRL2010}.

\begin{figure*}
    \begin{center}
        \includegraphics[scale=0.20]{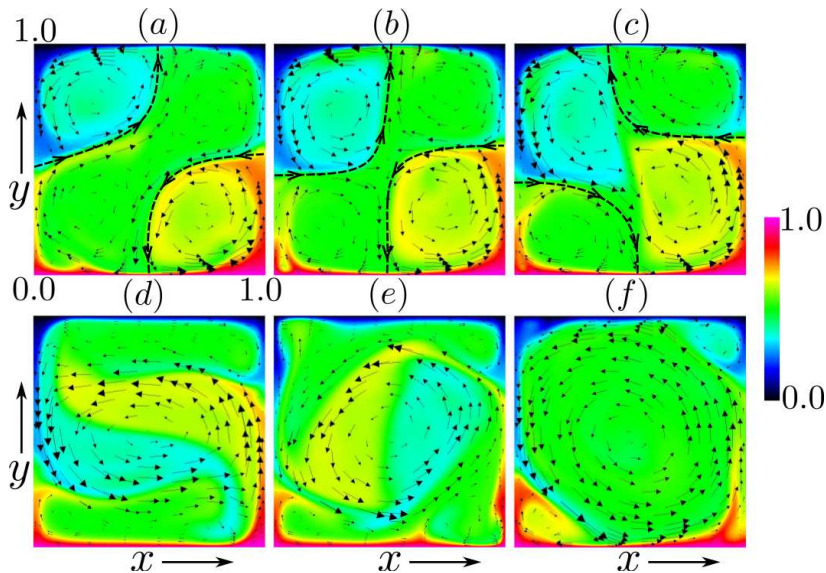}
    \end{center}
    \caption{Velocity and temperature profiles for six snapshots during the reversal: (a,b) Growth of corner rolls; (c) The two corner rolls at the upper-left and  bottom-right corners reconnect to form a large single roll.  The streamlines, represented by black curves, combine via vortex reconnections; (d,e) The flow reconfigures itself via rotation of the large roll formed after the reconnection.  There is a strong nonlinear interaction among the modes (1,1), (2,2), (1,3), and (3,1)  during these events; (f) The flow stabilizes to a quasi-steady state configuration.}
    \label{fig:reconnection}
\end{figure*}
Now we probe in detail the process of flow reversal by studying the flow structures of six snapshots during one of the reversals (see Fig.~\ref{fig:reconnection}).  A movie of the flow reversal can be downloaded from~\cite{movie}. The starting point of the reversal is the stable configuration shown in Fig.~\ref{fig:phase_transformation}(c).  At first, the mode (2,2)  (or the corner rolls) grows at the expense of the mode (1,1), as evident from the subfigures~\ref{fig:reconnection}(a,b).  The top-left and bottom-right corner rolls have vorticity in the same direction, hence they attract each other and come close due to the vortex dynamics in 2D~\cite{Tsinober:book}. Since the velocities of the streamlines are directed in opposite directions, they reconnect, thus converting two corner rolls into a single roll (see Fig.~\ref{fig:reconnection}(c)).   As a result of the reconnection, the new large roll has the cumulative vorticity in the same direction.
The reconnection event leads to a change in the flow topology, and  is similar to 2D magnetic field reconnections in magnetohydrodynamics~\cite{Priest:2000}.   The new large roll aligned along the -45 degree diagonal~(Fig.~\ref{fig:reconnection}(c)) has vorticity opposite to that of the large roll before the reversal (Fig.~\ref{fig:phase_transformation}(c))

\begin{figure}
    \begin{center}
       \includegraphics[scale=.4]{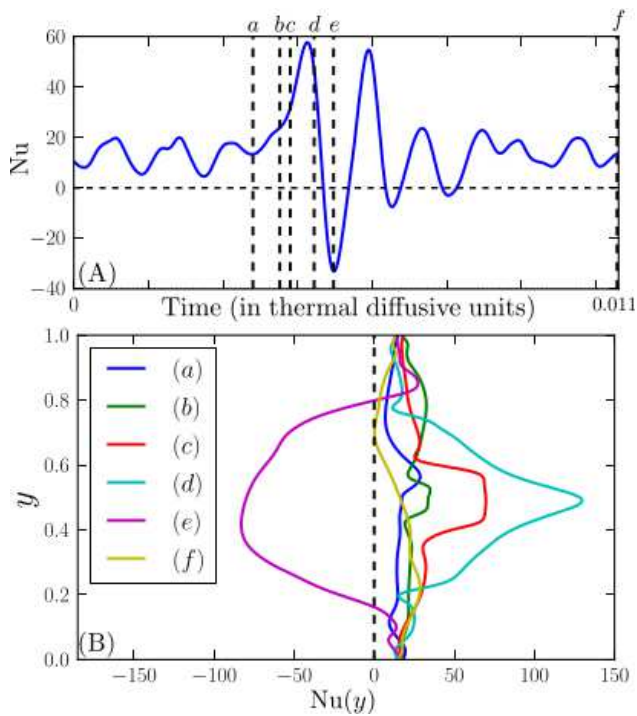}
   \end{center}
  \caption{(A) Time series of the global Nusselt number.  The six snapshots of Fig.~\ref{fig:reconnection} are marked as (a-f) in this plot.   (B) Plot of $\mathrm{Nu}(y)$ (the normalized heat transport for the cross-section at height $y$) vs.~$y$ during the reversal.  The six curves represent  the six snapshots of Fig.~\ref{fig:reconnection}.   The flow reconfigurations during the reversal lead to large fluctuations in $\mathrm{Nu}(y)$ in the bulk, ranging from strong positive values for (d) to strong negative values for (e).  Note however that the $\mathrm{Nu}(y)$ is positive near both the plates ($y \rightarrow 0,1$).}
  \label{fig:Nusselt_number_z}
\end{figure}

The strength of the new roll grows along with an emergence of two new secondary modes (1,3) and (3,1), which are generated as a result of triad interaction with the condition ${\bf k=p+q}$~\cite{Chandra:PRE2011}.  Note that $(1,3) = (2,2)+(-1,1)$ and $(3,1)=(2,2)+(1,-1)$.   The subsequent rotation of the newly formed roll leads to  flow configurations of Figs.~\ref{fig:phase_transformation}(d) and (e), which are dominated by  (1,1), (2,2), (1,3) and (3,1) modes.  Fig.~\ref{fig:reconnection}(d) illustrates three horizontally stacked rolls corresponding to the (1,3) mode.  After around 0.01 thermal diffusive time units, the system attains a quasi-steady state with the roll aligned along $-45$ degree diagonal with an opposite vorticity (Fig,~\ref{fig:reconnection}(f)) compared to the original one (Fig.~\ref{fig:phase_transformation}(c)).     The process would repeat for the next reversal with a difference that the bottom-left and the top-right corner rolls would reconnect during the next reversal.   

The modes (1,1) and (2,2) are the most dominant ones for the quasi-steady configurations, with the (1,1) mode switching sign between reversals.  The sign of the (2,2) mode or the sense of rotation of the corner rolls however remains unchanged after the reversal, as shown through symmetry arguments~\cite{Chandra:PRE2011}.  Another surprising observation is that the the amplitude of the (2,2) mode is always positive.  As a result, the hot plume of the corner rolls (at the bottom plate) always ascends via the vertical walls, and the cold plume descends via the vertical walls (see Figs.~\ref{fig:phase_transformation}(c) and \ref{fig:reconnection}(f)).   This seems to be a generic feature of many convection simulations and experiments, but its mathematical justification eludes us at present.    

We now discuss heat transport during the reversal.    The global Nusselt number $\mathrm{Nu} = \int [(d\bar{T}(y)/dy) + v  T ] d{\bf r}$ exhibits large fluctuations including negative values during a short time interval (see Fig.~\ref{fig:Nusselt_number_z}(A)).  Here, the nondimensionalized $\bar{T}(y)$ is the averaged temperature over the cross-section at height $y$.  The negative Nusselt number occurs for the flow configurations  resembling Fig.~\ref{fig:reconnection}(e) in which the hot fluid parcel in the middle of the box descends while the cold parcel ascends, contrary to generic situations when the hot parcels ascend and the cold ones descend.  Negative $\mathrm{Nu}$  may appear contradictory, but it could be understood in terms of   the heat flux through the cross-section at height $y$,  $ \mathrm{Nu}(y) =  d\bar{T}(y)/dy + \int_x v(x,y) T(x,y) dx$; here $\int_x$ is the integral over the line at height $y$.  We compute this quantity for various horizontal cross-sections for the flow configurations corresponding to Fig.~\ref{fig:reconnection}~(a-f).  As illustrated in Fig.~\ref{fig:Nusselt_number_z}(B), $\mathrm{Nu}(y)$ fluctuates significantly in the bulk for flow configurations (c,d,e) during the reversals, with strong positive values for (d), to strong negative values for (e).  Note however that $\mathrm{Nu}(y)$ is positive near the top and bottom plates for all cases.  The large fluctuations in $\mathrm{Nu}(y)$ in the bulk is a result of the rotation of the central role formed after reconnection.  We remark that the negative $\mathrm{Nu}$ for the two-dimensional convective flow is due to strong geometrical constraints faced by the rolls during the reversals, and may not be present in cylindrical convection. 

In summary, we show that flow reversals are caused by vortex reconnections of two attracting rolls.  The flow reconfigurations during the reversals are due to the nonlinear interactions among the large-scale modes (1,1), (2,2), (1,3), and (3,1).  We find large fluctuations in   heat transport during the reversals, which are due to the above restructuring of the flow.  The (2,2) mode, critical for the dynamics of flow reversals, is born after $\mathrm{Ra}=2\times 10^7$ (for $\mathrm{Pr}=1$), and its strength relative to the (1,1) mode decreases monotonically afterwards.  This is the reason why  flow reversals are observed only for a range of parameter values.   

The role of the large-scale modes described in the present letter is analogous to the ``cessation-led reversal" observed in turbulent convection in cylinder~\cite{Mishra:JFM2011} as well as in dynamo reversals~\cite{Gallet:GAFD2012}, where the quadrupolar mode (equivalent to (2,2) mode) dominates the dipolar mode (equivalent to (1,1) mode) during the reversal.  Future work for different geometries, and Prandtl and Rayleigh numbers would provide valuable insights that will help us build a comprehensive theory of reversals in convection and dynamo.

\acknowledgments{We are grateful to Paul Fischer and other developers of Nek5000 for opensourcing Nek5000  as well for providing valuable assistance during our work.  We thank Annick Pouquet,  Stephan Fauve, and J\"{o}rg Schumacher for very useful discussions. We also thank the Centre for Development of Advanced Computing  (CDAC) and the Computer Center of IIT Kanpur for providing us computing time.  Part of this work was supported by Swarnajayanti fellowship to MKV, and BRNS grant BRNS/PHY/20090310.}

%\newpage
%\begin{figure}
 %   \begin{center}
%        \includegraphics[scale=0.7]{fig_modes.pdf}
%    \end{center}
%    \caption{(a) Times series of the modes $(1,1)$, $(2,2)$, $(1,3)$ and
%    $(3,1)$. The modes $(1,1)$ and $(2,2)$ form two sets of triads, one with
%    $(1,3)$ and another with $(3,1)$ separately. Modes within a triad interact
  %  through nonlinear interactions. (b) Time series of the Nusselt number. The
%    dotted lines refer to the frames shown in fig.~(3). The Nusselt number shows
%    large fluctuations when the modes $(1,3)$ and $(3,1)$ become dominant.}
%    \label{fig:mode_timeseries}
% \end{figure}

\end{document}